\documentclass[a4paper,clock]{article}
\usepackage[dvips]{epsfig}
\usepackage{amssymb}
\usepackage{bm}
\usepackage{dcolumn}

\catcode`\@=11
\def\marginnote#1{}
\newcount\hour
\newcount\minute
\newtoks\amorpm
\hour=\time\divide\hour by60
\minute=\time{\multiply\hour by60 \global\advance\minute
by-\hour}\edef\standardtime{{\ifnum\hour<12
\global\amorpm={am}%
        \else\global\amorpm={pm}\advance\hour by-12 \fi
        \ifnum\hour=0 \hour=12 \fi
        \number\hour:\ifnum\minute<10
0\fi\number\minute\the\amorpm}}
\edef\militarytime{\number\hour:\ifnum\minute<10
0\fi\number\minute}

\def\draftlabel#1{{\@bsphack\if@filesw {\let\thepage\relax
   \xdef\@gtempa{\write\@auxout{\string
      \newlabel{#1}{{\@currentlabel}{\thepage}}}}}\@gtempa
   \if@nobreak \ifvmode\nobreak\fi\fi\fi\@esphack}
        \gdef\@eqnlabel{#1}}
\def\@eqnlabel{}
\def\@vacuum{}
\def\draftmarginnote#1{\marginpar{\raggedright\scriptsize\tt#1}}
\def\draft{\oddsidemargin -.5truein
        \def\@oddfoot{\sl preliminary draft \hfil
        \rm\thepage\hfil\sl\today\quad\militarytime}
        \let\@evenfoot\@oddfoot \overfullrule 3pt
        \let\label=\draftlabel
        \let\marginnote=\draftmarginnote

\def\@eqnnum{(\theequation)\rlap{\kern\marginparsep\tt\@eqnlabel}%
\global\let\@eqnlabel\@vacuum}  }

\def\numberbysection{\@addtoreset{equation}{section}
        \def\theequation{\thesection.\arabic{equation}}}

\def\underline#1{\relax\ifmmode\@@underline#1\else
 $\@@underline{\hbox{#1}}$\relax\fi}

\catcode`@=12
\relax

\numberbysection

\advance \topmargin by -\headheight
\advance \topmargin by -\headsep

\evensidemargin \oddsidemargin
\def\nonu{\nonumber}
\def\br{\begin{eqnarray}}
\def\er{\end{eqnarray}}
\def\be{\begin{equation}}
\def\ee{\end{equation}}

\def\({\left(}
\def\){\right)}

\relax

\def\pa{\partial}

\def\tp0{\Theta_{+}^{(0)}}
\def\tm0{\Theta_{-}^{(0)}}

\def\f#1#2#3 {f^{#1#2}_{#3}}

\def\win1{{\sf w_{1+\infty}}}

\def\Win1{{\sf W_{1+\infty}}}

\def\rlx{\relax\leavevmode}
\def\inbar{\vrule height1.5ex width.4pt depth0pt}
\def\IZ{\rlx\hbox{\sf Z\kern-.4em Z}}
\def\IR{\rlx\hbox{\rm I\kern-.18em R}}
\def\IC{\rlx\hbox{\,$\inbar\kern-.3em{\rm C}$}}
\def\IN{\rlx\hbox{\rm I\kern-.18em N}}
\def\IO{\rlx\hbox{\,$\inbar\kern-.3em{\rm O}$}}
\def\IP{\rlx\hbox{\rm I\kern-.18em P}}
\def\IQ{\rlx\hbox{\,$\inbar\kern-.3em{\rm Q}$}}
\def\IF{\rlx\hbox{\rm I\kern-.18em F}}
\def\IG{\rlx\hbox{\,$\inbar\kern-.3em{\rm G}$}}
\def\IH{\rlx\hbox{\rm I\kern-.18em H}}
\def\II{\rlx\hbox{\rm I\kern-.18em I}}
\def\IK{\rlx\hbox{\rm I\kern-.18em K}}
\def\IL{\rlx\hbox{\rm I\kern-.18em L}}
\def\one{\hbox{{1}\kern-.25em\hbox{l}}}
\def\0#1{\relax\ifmmode\mathaccent"7017{#1}%
B        \else\accent23#1\relax\fi}

\def\RQE#1#2#3{{\sl Radiophysics and Quantum Electronics} {\bf #1} (#2) #3}
\def\PRL#1#2#3{{\sl Phys. Rev. Lett.} {\bf#1} (#2) #3}
\def\NPB#1#2#3{{\sl Nucl. Phys.} {\bf B#1} (#2) #3}

\def\PRE#1#2#3{{\sl Phys. Rev.} {\bf E#1} (#2) #3}
\def\PRA#1#2#3{{\sl Phys. Rev.} {\bf A#1} (#2) #3}

\def\PLA#1#2#3{{\sl Phys. Lett.} {\bf #1A} (#2) #3}

\def\PLB#1#2#3{{\sl Phys. Lett.} {\bf #1B} (#2) #3}

\def\PTP#1#2#3{{\sl Prog. Theor. Phys.} {\bf #1} (#2) #3}

\def\IJMPA#1#2#3{{\sl Int. J. Mod. Phys.} {\bf A#1} (#2) #3}
\def\IJMPB#1#2#3{{\sl Int. J. Mod. Phys.} {\bf B#1} (#2) #3}

\def\JPA#1#2#3{{\sl J. Physics} {\bf A#1} (#2) #3}

\def\JHEP#1#2#3{{\sl JHEP} {\bf #1} (#2) #3}

\def\SAM#1#2#3{{\sl Stud. Appl. Math.} {\bf #1} (#2) #3}

\def\OL#1#2#3{{\sl Optics Letters} {\bf #1} (#2) #3}
\def\Nonl#1#2#3{{\sl Nonlinearity} {\bf #1} (#2) #3}

\def\JPAMG#1#2#3{{\sl J. Physics A: Math. Gen.} {\bf A#1} (#2) #3}
\def\JGP#1#2#3{{\sl Journal of Geometry and Physics} {\bf #1} (#2) #3}
\def\EPL#1#2#3{{\sl Europhysics Letters} {\bf #1} (#2) #3}
\def\ScR#1#2#3{{\sl Sci. Rep.} {\bf #1} (#2) #3}

\def\JPCS#1#2#3{{\sl J. Phys.: Conf. Ser} {\bf #1} (#2) #3}

\def\AML#1#2#3{{\sl Applied Mathematics Letters} {\bf #1} (#2) #3}

\def\ND#1#2#3{{\sl  Nonlinear Dyn. } {\bf #1} (#2) #3}
\def\RF#1#2#3{{\sl Results in Physics} {\bf #1} (#2) #3}
\def\AMP#1#2#3{{\sl Anal. Math. Phys.} {\bf #1} (#2) #3}
\def\JMAA#1#2#3{{\sl Journal of Mathematical Analysis and Applications} {\bf #1} (#2) #3}
\def\CTP#1#2#3{{\sl Commun. Theor. Phys. } {\bf #1} (#2) #3}
 
\begin{document}

\begin{titlepage}

\vspace{.2in}
\begin{center}
{\large\bf Modified AKNS model, Riccati-type pseudo-potential approach and infinite towers of quasi-conservation laws}
\end{center}

\vspace{.2in}

\begin{center}

H. Blas$^{(a)}$,  M. Cerna Magui\~na$^{(b)}$ and  L.F. dos Santos$^{(c)}$

\par \vskip .2in \noindent

$^{(a)}$Instituto de F\'{\i}sica\\
Universidade Federal de Mato Grosso\\
Av. Fernando Correa, $N^{0}$ \, 2367\\
Bairro Boa Esperan\c ca, Cep 78060-900, Cuiab\'a - MT - Brazil. \\
$^{(b)}$ Departamento de Matem\'atica\\
Universidad Nacional Santiago Ant\'unez de Mayolo\\
Campus Shancay\'an, Av. Centenario 200, Huaraz - Per\'u\\
$^{(c)}$ Centro Federado de Educa\c c\~ao Tecnologica-CEFET-RJ\\
Campus Angra dos Reis, Rua do Areal, 522, Angra dos Reis- RJ -Brazil

\normalsize
\end{center}

\vspace{.3in}

\begin{abstract}
\vspace{.3in}

A dual Riccati-type pseudo-potential formulation is introduced for a modified AKNS system (MAKNS) and infinite towers of novel anomalous conservation laws are uncovered. In addition, infinite towers of exact non-local conservation laws are uncovered in a linear formulation of the system. It is shown that certain modifications of the  non-linear Schr\"odinger model (MNLS) can be obtained through a reduction process starting from the MAKNS model. So, the novel infinite sets of quasi-conservation laws and related anomalous charges are constructed by an unified and rigorous approach based on the Riccati-type pseudo-potential method, for the standard NLS and modified MNLS cases, respectively. The non-local properties, the complete list of towers of infinite number of anomalous charges and the (non-local) exact conservation laws of the quasi-integrable systems, such as the deformed Bullough-Dodd, Toda, KdV and SUSY sine-Gordon systems can be studied in the framework presented in this paper. Our results may find many applications since the AKNS-type system arises in several branches of non-linear physics, such as Bose-Einstein condensation, superconductivity and soliton turbulence.

\end{abstract}

\end{titlepage}

\section{Introduction}

Certain non-linear field theory models with relevant physical applications and modeling solitary waves are not integrable. Recently, the concept of quasi-integrability for some deformations of integrable models has been put forward. In that context, some properties of the deformations of the soliton models, such as SG, NLS, Toda, KdV, Boullogh-Dodd and SUSY-SG have been examined in the frameworks of the anomalous zero-curvature formulations \cite{jhep1, jhep2, jhep6, jhep3, toda, cnsns, npb, jhep4, jhep5, epl} and the deformations of the Riccati-type pseudo-potential approach \cite{npb1, jhep33}. 
  
The quasi-integrable models set forward in the literature \cite{jhep1, jhep2,jhep3,jhep4, jhep5, jhep6, npb, jhep33} possess important structures, such as infinite sets of non-local conserved charges, and some types of linear formulations. In this context, the Riccati-type representations have recently been presented for the deformed KdV and sine-Gordon models \cite{jhep33, npb1}.  Moreover, in the context of the Riccati-type pseudo-potential approach to quasi-integrability, it has been shown that the deformed SG and KdV models \cite{npb1, jhep33} can be formulated as the compatibility condition of certain linear systems of equations and that they possess infinite towers of exact non-local conservation laws.   

In this paper we will tackle the problem of extending the Riccati-type pseudo-potential formalism, which has been used for a variety of well known integrable systems, to deformed AKNS models. The new properties mentioned above have been examined for the deformations of the relativistic invariant sine-Gordon model with topological solitons and the non-relativistic KdV model with non-topological and unidirectional solitons, respectively. Conventionally, both of  them are defined for real scalar fields. So, it would be desirable to examine those properties for NLS-type models defined for a complex field with envelope solitons. Some of the mentioned properties have recently been examined  for a deformed NLS model in \cite{paper1} by direct construction of novel quasi-conservation laws starting from the eqs. of motion.    

In general, those integrable models can be formulated in the framework of  the AKNS system, from which they can be obtained through relevant reduction processes. The NLS-type model stands in the same level of importance as the KdV-type and SG-type  models in their potential applications, since they are ubiquitous in all areas of nonlinear physics, such as  Bose-Einsten condensation and superconductivity \cite{be1, frantzeskakis, tanaka1}, soliton gas and soliton turbulence in fluid dynamics \cite{gauss, pla1, turbu, prlgas}, the Alice-Bob physics \cite{alice}, and the understanding of a kind of triality among the gauge theories, integrable models and gravity theories (see \cite{nian} and references therein).    

We perform a particular deformation of the Riccati-type pseudo-potential approach related to the AKNS system \cite{nucci},  from which the NLS-like modified model is obtained through a particular reduction process. A modified  AKNS system (MAKNS) is defined by introducing a deformed potential $V$ and some auxiliary fields into the pair of Riccati-type system of equations and another system of equations for the set of auxiliary fields, such that the compatibility condition applied to the extended system gives rise to the modified AKNS model equations of motions. Then, it is constructed a set of infinite number of quasi-conservation laws order by order in powers of the spectral parameter. 

Remarkably, a new dual system of the Riccati-type pseudo-potential approach is introduced which provides infinite sets of novel anomalous conservation laws. Those quasi-conservation laws encompass, as a subset, the ones obtained by a direct constructive method for the MNLS model in the companion paper \cite{paper1}. So, in this paper it will be shown unified and rigorous constructions of the novel anomalous charges for  the standard NLS and modified MNLS cases, respectively.

In addition, in the framework  of the pseudo-potential approach, it is  proposed a linear system of equations whose compatibility condition gives rise to the MAKNS equations of motion. As an application of the linear system formulation of the modified AKNS model, it is obtained a pair of infinite towers of exact non-local conservation laws.  A particular reduction $MAKNS \rightarrow MNLS$ allows one to reproduce the relevant quantities of the MNLS model out of the ones constructed for  the MAKNS system.   
    
This paper is organized as follows: The section \ref{sec:riccati} considers a particular deformation of the $sl(2)$ AKNS model in the context of the Riccati-type pseudo-potential approach, and it discusses a particular reduction to the modified NLS model. An infinite set of quasi-conservation laws are constructed. In sec. \ref{sec:dual1} a dual Riccati-type formulation and novel anomalous charges are presented. 
In sec. \ref{linear} it is found a linear system formulation of the deformed AKNS model and constructed an infinite set of non-local conservation laws. In sec. \ref{ap:conclu} we present our conclusions and discussions. The appendices \ref{fsca1} , \ref{ap:chi} and \ref{app:uchid} present the components of the expansions in power of $\zeta$ of the Riccati-type pseudo-potentials. 

\section{Riccati-type pseudo-potential and modified $sl(2)$ AKNS model}
 \label{sec:riccati}

The standard NLS model can be  obtained as a special reduction of the AKNS system; so, in the next sections we consider a convenient deformation of the usual pseudo-potential approach to the AKNS  integrable field theory. Subsequently, we will discuss its reduction process leading to the modified NLS model. In \cite{nucci} it has been generated the both Lax equations and B\"acklund transformations 
for well-known non-linear evolution equations using the concept of pseudo-potentials and the related  
properties of the Riccati equation. These applications have been done in the context of a variety of  integrable systems (sine-Gordon, KdV, NLS, etc), and allow the Lax pair formulation, the construction of  conservation laws and the B\"acklund transformations for them \cite{nucci, prl1}. 

So, let us consider the system of Riccati-type equations 
\br
\label{ricc1}
\pa_x u &=& -2 i \zeta \, u + q + \, \bar{q} \,\, u^2,\\
\label{ricc2}
\pa_t u &=&  2 A\, u - C\, u^2 + B + r - u\, s,
\er
where $u$ is the Riccati-type pseudo-potential, $r$ and  $s$ are auxiliary fields, and $q$ and $\bar{q}$ are the fields of the model. Let us assume
\br
\label{A11}
A & \equiv & - 2 i \zeta^2 -  \frac{1}{2} i\, V^{(1)},\\
\label{A22}
B & \equiv & 2 \zeta q + i \pa_x q,\\
\label{A33}
C & \equiv & -2 \zeta \bar{q} + i \pa_x \bar{q},\,\,\,\,\,V^{(1)} \equiv  \frac{d V[\rho]}{d \rho},\,\,\,\rho \equiv \bar{q} q, 
\er
where $V(\bar{q} q)$ is the potential of the modified AKNS equation (MAKNS) and $\zeta$ is the spectral parameter. In fact, the form of (\ref{ricc1}) is similar to the first Riccati equation for the standard AKNS model. The functions $B$ and $C$ are the same as in the usual second Riccati equation; whereas, the funtion $A$ has been modified to contain a generalized  potential as compared to the standard  AKNS model \cite{prl1}.

We consider the following equations for the auxiliary fields $r(x,t)$ and  $s(x,t)$
\br
\label{ricc1r}
\pa_x r &=& q \, s +  (-2 i \zeta + Q \, u)\, r,\\
\label{ricc2s}
\pa_x s &=& Q \,r - 2  \, \bar{q}\, r + u \bar{q}\, s + 2 i X,\\
X & \equiv & - \pa_x  \( \frac{1}{2} V^{(1)} + \bar{q} q \),  \label{Xanom}
\er
where $Q$ is an arbitrary field. So, one has a set of two deformed Riccati-type equations for the pseudo-potential $u$ (\ref{ricc1})-(\ref{ricc2}) and a system of equations (\ref{ricc1r})-(\ref{ricc2s}) for the auxiliary fields $r$ and  $s$. 

Notice that, for the integrable AKNS model one has the potential 
\br
\label{nlspot}
V_{NLS}(\bar{q} q) = - \( \bar{q} q \)^2\, \rightarrow \, V_{NLS}^{(1)}(\bar{q} q) = - 2(\bar{q} q),
\er
and, therefore,  $X=0$ in (\ref{Xanom}), and so the auxiliary system of  eqs. (\ref{ricc1r})-(\ref{ricc2s}) possesses the trivial solution $r=s=0$. Inserting this trivial solution into the system  (\ref{ricc1})-(\ref{ricc2}) and considering  the potential (\ref{nlspot}), one has a set of two Riccati equations for the standard AKNS model and they  play an important role in order to study its properties, such as the derivation of the infinite number of conserved charges and the B\"acklund transformations, relating the fields $(q ,\bar{q})$ with another set of solutions $(q',\bar{q}')$ \cite{prl1}.   
  
Note that only the $t-$component $\pa_t u$ of the Riccati equation associated to the ordinary AKNS equation has been  deformed away from the AKNS potential (\ref{nlspot}), and it carries all the information regarding the deformation of the model which are encoded in the potential $V(\bar{q} q )$ and the auxiliary fields $r(x,t)$ and  $s(x,t)$. The form of the $x-$component $\pa_x u $ remains the same as the usual Riccati equation associated to the AKNS model. 
 
We have computed the compatibility condition   $[\pa_{t}\pa_x  u - \pa_x \pa_{t} u ]=0$ for the Riccati-type equations (\ref{ricc1})-(\ref{ricc2}), taking into account  the auxiliary system of equations (\ref{ricc1r})-(\ref{ricc2s}) and then derived  the eqs. of motion for the fields $q $ and $\bar{q}$
\br
\label{mnls11}
i \pa_t q + \pa^2_x q - V^{(1)} q &=& 0,\\
-i \pa_t \bar{q} + \pa^2_x \bar{q} - V^{(1)} \bar{q} &=& 0.\label{mnls22}
\er
This is a modified $sl(2)$ AKNS system (MAKNS) for arbitrary potential of type $V(\bar{q} q)$. An important observation in the constructions above is that $\frac{\pa}{\pa t} \zeta =0$, as it can be checked by direct computation using 
 the system (\ref{ricc1})-(\ref{ricc2}) and (\ref{ricc1r})-(\ref{ricc2s}), provided that the system of eqs. (\ref{mnls11})-(\ref{mnls22}) is satisfied. So, the modified system MAKNS possesses an isospectral parameter $ \zeta $. 
 
Next, providing that the identifications  
\br
\label{nlspsi}
q \equiv  i (-\eta)^{1/2}\, \psi,\,\,\, \bar{q} \equiv  -  i (-\eta)^{1/2} \, \bar{\psi},\,\,\,\,\eta \in \IR
\er 
are performed in the system (\ref{mnls11})-(\ref{mnls22}), where $\bar{\psi}$ stands for complex conjugation of the field $\psi$, one can obtain the modified  NLS model
\br
\label{nlsd}
i \frac{\pa}{\pa t} \psi(x,t) + \frac{\pa^2}{\pa x^2} \psi(x,t) -  V^{(1)}(|\psi(x,t)|^2)\psi(x,t) &=&  0,\\
V^{(n)}(I)&\equiv &\frac{d^n}{d I^n} V(I),\,\,\,\, I \equiv \bar{\psi} \psi, \er 
where $\psi$ is a complex scalar field. If the potential and its derivatives are taken as 
\br \label{potder}
V(I)= - \eta I^2,\,\,\,V^{(1)}(I) =- 2\eta I \,\,\, \mbox{and}\,\,\,\, V^{(2)}(I)= - 2\eta.
\er
the reduction (\ref{nlspsi}) defines the standard NLS model. This is a process, we have just mentioned above, through which the standard NLS model is obtained as a special reduction of the AKNS system.

Let us emphasize that for the standard NLS model we have the  trivial solution of the system  (\ref{ricc1r})-(\ref{ricc2s}), i.e. $X=0 \rightarrow r=s=0$, and the existence of the Lax pair of de ordinary NLS model reflects in its equivalent Riccati-type representation, provided by the system (\ref{ricc1})-(\ref{ricc2}) with the well known potential (\ref{potder}) \cite{nucci, prl1}. 

Next, let us consider a special space-time symmetry related to soliton-type solutions of the model. So, consider a reflection around a fixed point $(x_{\Delta},t_{\Delta})$
\br
\label{par1}
\widetilde{{\cal P}}:  (\widetilde{x},\widetilde{t}) \rightarrow (-\widetilde{x},-\widetilde{t});\,\,\,\,\,\,\,\,\widetilde{x} = x - x_{\Delta},\,\,\widetilde{t} = t- t_{\Delta}. 
\er 
The transformation $\widetilde{{\cal P}}$ defines a shifted parity ${\cal P}_{s}$ for the spatial variable $x$  and a delayed time reversal ${\cal T}_d$ for the time variable $t$. Notice that, when $x_{\Delta}=0$ ($t_\Delta=0$), ${\cal P}_{s}$ (${\cal T}_d$) is reduced to the pure parity ${\cal P}$ (pure time reversal ${\cal T}$).  

We define the quasi-integrable MAKNS model for field configurations $q$ and $\bar{q}$ satisfying (\ref{mnls11})-(\ref{mnls22}) such that the fields and the deformed potential transform under the space-time transformation (\ref{par1}) as
\br 
\label{aknsparity}
\widetilde{{\cal P}}(q)  = \bar{q},\,\,\,\,\,\, \widetilde{{\cal P}}(\bar{q})  = q,\,\,\,\,\,\mbox{and } \,\,\,\,\,\widetilde{{\cal P}}[ V(\rho) ] = V(\rho),\,\,\,\rho \equiv \bar{q}  q.
\er
Under this transformation one has that  $X$  from (\ref{Xanom}) becomes an odd function
\br
\label{Xtr}
\widetilde{{\cal P}}( X ) =  - X.
\er
Next, let us discuss the relevant conservation laws in the context of the Riccati-type system (\ref{ricc1})-(\ref{ricc2}) and the auxiliary equations (\ref{ricc1r})-(\ref{ricc2s}). So, substituting the expression for $u^2$ from (\ref{ricc1}) into (\ref{ricc2}) and considering (\ref{ricc1r})-(\ref{ricc2s}), one gets the following relationship
\br
\label{qcons}
\pa_t [ i \bar{q} \,u  ] - \pa_x \Big[ 2 i \zeta \bar{q} \, u - \bar{q} q +  u \, \pa_x \bar{q} \Big] &=& i \bar{q} (r - s \, u). 
\er
Defining the r.h.s. of (\ref{qcons}) as
\br
\label{qcons11}
\chi \equiv i \bar{q} (r - s \, u),
\er 
and using the system  (\ref{ricc1})-(\ref{ricc2}) and the auxiliary equations (\ref{ricc1r})-(\ref{ricc2s}) one can write a first order differential equation for the auxiliary field $\chi$ 
\br
\label{chi0}
\pa_x \chi &=& \( -2 i \zeta + 2 u \bar{q} + \frac{\pa_ x \bar{q}}{\bar{q}}\) \chi + 2 \bar{q}\, u X.
\er
The eqs. (\ref{qcons}) and (\ref{chi0}) will be used below in order to uncover an infinite tower of quasi-conservation laws associated to the modified AKNS model (\ref{mnls11})-(\ref{mnls22}).  We will construct  the relevant charges order by order in powers of the parameter $\zeta$. So,  let us consider the expansions
\br
\label{expan}
 u = \sum_{n=1}^{\infty} u_n \, \zeta^{-n},\,\,\,\,  \chi= \sum_{n=1}^{\infty} \chi_n  \zeta^{-n-1}.
\er
The coefficients $u_n$ of the expansion above  can be determined order by order in powers of $\zeta$ from  the Riccati equation (\ref{ricc1}).  In appendix \ref{fsca1} we provide the recursion relation for the $u_n\, 's$ and the expressions for the first $u_n$. Likewise, using the results for the $u_n\, 's$ we get the relevant expressions for the  $\chi_n\, 's$ from (\ref{chi0}). The first components $\chi_n$ are provided in appendix \ref{ap:chi}.

Then, making use of the $u_n$ and $\chi_n$ components of the expansions of $u$ and $\chi$, respectively, provided in (\ref{expan}), one can find the conservation laws, order by order in powers of $\zeta$. So, by inserting those expansions into  the eq. (\ref{qcons}) one has that  the coefficient of the $n'$th order term becomes
\br
\label{anocons}
\pa_{t} a_{x}^{(n)} &+& \pa_x a_{t}^{(n)} = \chi_{n-1},\,\,\,\,\,n=0,1,2,3,....;\, \chi_{0}\equiv 0\\
a_{x}^{(n)} &\equiv & i \bar{q}\, u_n ,\,\,\,\,a_{t}^{(n)} \equiv   -\( 2 i \bar{q} \, u_{n+1} - \bar{q} q \, \delta_{0, n}  + \pa_x \bar{q} \, u_{n}  \),\,\,\,\,\,u_0 \equiv 0.
\er
Notice that making the substitution $\chi_{n-1} \equiv 0$ into the eq. (\ref{anocons}) one can get the tower of exact conservation laws of the usual AKNS system. A truly conservation law character of this equation, at each order $n$, remains to be clarified, since the field components $\chi_{n-1}$ in the r.h.s. of (\ref{anocons}), as they can be seen in the appendix \ref{ap:chi}, do not present the adequate forms to be directly incorporated into  the l.h.s. of the conservation laws. We will tackle this construction order by order for each $\chi_{n-1}$ component. Notice that analogous quasi-conservation laws have been obtained in the context of the anomalous zero-curvature formulation of the modified NLS model and its associated anomalous Lax pair in \cite{jhep3}.  

We will show below that the r.h.s. of (\ref{anocons}) for  $\chi_1, \chi_2$ and $\chi_3$ can be written in the form $ \chi_j \equiv \pa_{x} \chi^x_j + \pa_{t} \chi^t_j$, with $\chi^x_j$ and $\chi^t_j$ being certain local functions of $\{\bar{q}, q, V\}$ and their $x$ and  $t-$derivatives; i.e. there exist local expressions for some $\chi_j\, (j=1,2,3)$, such that the eq. (\ref{anocons}) provides a proper local conservation law. 

Let us compute the charges order by order in $n$ using the eq. (\ref{anocons}) and the relevant expressions presented in the appendices \ref{fsca1} and \ref{ap:chi}.  

The  {\bf zero'th} order provides a trivial identity.
 
{\bf The order $n=1$ and the field normalization}

In this case the anomaly is trivial $\chi_0 =0$. So, one has
\br
\label{n1}
\pa_t \( \frac{1}{2} \bar{q} q \) - \pa_x \( \frac{1}{2} i \bar{q} \pa_x q -  \frac{1}{2} i q \pa_x \bar{q}   \) =0.
\er
It provides the conserved charge
\br
\label{n1nor}
N = \int dx \, \bar{q} q
\er

{\bf The order $n=2$ and momentum conservation}

At this order one has 
\br
\label{n2}
\pa_t \( \frac{1}{4} i \bar{q} \pa_x q\) + \pa_x \( \frac{1}{4} [ (\bar{q} q)^2 + \bar{q} \pa_x^2 q - \pa_x \bar{q}\pa_x q]\) &=&\chi_1.\er
The function  $\chi_1$ can be rewritten as
\br
\label{chi1}
\chi_1 &=& \frac{1}{2} \pa_x [ F(\rho) ],\,\,\,\,\rho \equiv \bar{q} q . \\
\label{fv}
F(I) & \equiv &\frac{1}{2} \rho \frac{d}{d \rho}V(\rho)-\frac{1}{2} V(\rho)+\frac{1}{2} \rho^2. 
\er
So, from (\ref{n2}), taking into account  (\ref{chi1}), one can write the conserved charge
\br
\label{n2mo}
P =   i \int dx \, \( \bar{q} \pa_x q - q \pa_x \bar{q} \)
\er

{\bf The order $n=3$ and energy conservation}

One has the conservation law
\br
\label{n3}
\pa_t [ -\frac{1}{8} (\bar{q} q)^2 - \frac{1}{8} \bar{q} \pa^2_x q ] -\pa_x \(2 i \bar{q} u_4 + \pa_x \bar{q} u_3 \) &=& \chi_2.       
\er
The function $\chi_2$ can be rewritten as
\br
\label{chixt}
 \chi_2 \equiv  -\frac{1}{8}  \pa_t V -\frac{1}{8} \pa_t (\bar{q} q)^2 - \frac{1}{4} i \pa_x\Big[ X \bar{q} q - X (q  \pa_x \bar{q} - \bar{q}  \pa_x q)\Big].
\er
So, (\ref{n3}) provides the conserved charge
\br
\label{energy}
H_{MNLS} =   \int dx \, [\, \pa_x \bar{q} \pa_x q + V( \bar{q} q)  \,].
\er
Notice that in order to get the identity (\ref{chixt}) we have used the eqs. of motion (\ref{mnls11})-(\ref{mnls22}).
Since we have considered $\chi_2 \neq 0$ in the r.h.s. of (\ref{n3}), which carries the effect of the modified potential,  the expression of the energy  (\ref{energy})  is valid for the general MNLS model.  In particular, for the ordinary AKNS the energy follows directly  from the l.h.s. of (\ref{n3}) (provided that $\chi_2 =0 $ in the r.h.s.  of that eq.), i.e.  $H_{NLS} =   \int dx \, [\, \pa_x \bar{q} \pa_x q + V_{NLS}( \bar{q} q)  \,]$, where $V_{NLS} = - (\bar{q} q )^2$ as in (\ref{nlspot}).

{\bf The order $n=4$:  A first trivial charge and its associated anomalous charge}

One has 
\br
\label{charge4}
\pa_t \(- \frac{3}{16} i \bar{q} q \bar{q} \pa_x q -\frac{1}{32} i \pa_x (\bar{q} q)^2  - \frac{1}{16} i \bar{q} \pa^3_x q \)- \pa_x [ 2 i \bar{q} \, u_5 + \pa_x \bar{q} \, u_4 ] &=& \chi_3.
\er
Remarkably, the expression for $\chi_3$ can be written as  
\br
\label{chi3xt}
\chi_3 & \equiv & \pa_x [ \chi^{(3)}_{x} ] + \pa_t [ \chi^{(3)}_{t} ],\\
   \chi^{(3)}_{t} & = & - \frac{3}{16} i \bar{q} q \bar{q} \pa_x q  - \frac{1}{16} i \bar{q} \pa^3_x q, \label{chi3t}\\
   \label{chi3x}
    \chi^{(3)}_{x} & = & \frac{3}{8} X \bar{q} \pa_x q + \frac{3}{8}  H_1(\rho) + \frac{1}{8} \bar{q} q  \pa_x X - \frac{1}{8} X \pa_{x} (\bar{q} q) - \frac{1}{16} [\pa_{x} (\bar{q} q) ]^2+ \frac{3}{8} \bar{q} q \pa_{x}\bar{q} \pa_x q- \frac{3}{8} H_2(\rho) + \nonumber\\ 
&& \frac{3}{32} i \bar{q} \pa_t q^2 - \frac{1}{16} V^{(1)} \( q \pa^2_x \bar{q} + \bar{q} \pa^2_x q  \) + \frac{1}{16}   V^{(1)}  \pa_x q \pa_x \bar{q} +\nonumber\\ 
&& \frac{1}{16} i [\pa_t q \pa^2_x \bar{q} - \pa_x \pa_t q \pa_x \bar{q} + \bar{q} \pa_t \pa^2_x q].  
    \\   
    \frac{d}{d\rho}H_1(\rho) & \equiv & -(V^{(2)}/2+1) \rho^2,\,\,\,\,\, \frac{d}{d\rho}H_2(\rho) \equiv \rho V^{(1)},\,\,\,\,\,\rho \equiv \bar{q}  q.
\er
Therefore, at this order, the eq. (\ref{charge4}) can be written as an exact conservation law. However, taking into account the term $\pa_t \chi^{(3)}_{t}$ of $\chi_3$ in (\ref{chi3xt})-(\ref{chi3t}) and the relevant terms in  the l.h.s. of (\ref{charge4}) with partial $t-$derivatives one gets a fourth order trivial charge $Q^{(4)}=0$, provided that the surface term $\sim \pa_x (\bar{q} q)^2$ is dropped, since upon integration in $x$ in order to define the charge it vanishes for suitable boundary conditions. So, at this order of the above formulation, the charge $Q^{(4)}$ trivially vanishes.

However,  at this order and in the higher order ones, one can define an asymptotically conserved charge for the MAKNS model 
\br
\label{charge4a}
Q^{(4)}_a =  \frac{i}{2} \int dx \, \Big[ 3 \bar{q}  q \( \bar{q} \pa_x q - q \pa_x \bar{q} \) + \bar{q} \pa^3_x q - q \pa^3_x \bar{q}  \Big],
\er
such that 
\br
\label{qa44}
\frac{d}{dt}  Q^{(4)}_a     &=& \hat{\tau}\\
\hat{\tau} &=& - 8 \int\, dx \, \chi_3,\label{qa441}\\
 &=& - 8 \int\, dx \, \Big\{ \frac{1}{8} \Big[  3 \pa_x( \bar{q} \pa_x q  X) - 3 \pa_x( \bar{q} \pa_x q) X +  3\bar{q} (\bar{q} q^2+\pa^2_x q) X + \bar{q} q \pa^2_x X \Big]\Big\} \nonumber\\
\label{qa442}
                 &=&  -  \int\, dx \, [ 3(\bar{q} q)^2 X + \bar{q} q \pa^2_x X - 3 \pa_x \bar{q} \pa_x q X],
\er
where in (\ref{qa441}) the expression of $ \chi_3$ from (\ref{chis}) must be inserted and the final form of the  anomaly density in (\ref{qa442} ) is obtained by dropping a surface term. Notice that the anomaly density in (\ref{qa442}) possesses an odd parity under (\ref{par1}) and (\ref{aknsparity}) taking into account that $X$ is an odd function according to (\ref{Xtr}). Therefore, one has $\int dt \int dx\, \chi_3 = 0$ implying the asymptotically conservation of the charge $Q^{(4)}_a$.   

The charge $Q^{(4)}_a $ in (\ref{charge4a}) takes the same form as the fourth order charge in the standard AKNS model. In fact, when the r.h.s. of (\ref{charge4}) vanishes, i.e. $\chi_3 =0$, one has a charge similar in form to the one in (\ref{charge4a}),  conveniently rewritten by discarding surface terms. Taking into account the reduction process (\ref{nlspsi}) one can get a similar anomalous charge for the MNLS model, as presented in  \cite{jhep3, jhep4, jhep5}. In fact, upon the reduction  (\ref{nlspsi})  the anomalous charge  $Q^{(4)}_a $ in (\ref{charge4a}) corresponds to the one for the MNLS model in sec. 3.5 of \cite{paper1}. 
 
{\bf The order $n=5$ and the quasi-conserved charge}

At this order one has
\br
\label{charge5}
 \frac{1}{32} \pa_t  \Big[ 2 (\bar{q} q)^3  + 5 \bar{q}^2 (\pa_x q)^2 + 6 \bar{q}q \( \pa_x q \pa_x \bar{q} + \bar{q}  \pa^2_x q\) + \bar{q} q^2 \pa^2_x \bar{q} + \bar{q} \pa^4_x q  \Big]- \pa_x [ 2 i \bar{q} \, u_6 + \pa_x \bar{q} \, u_5 ] = \chi_4.
\er
Likewise, the expression for $\chi_4$ can be written as  
\br
\label{chi4xt}
\chi_4 & \equiv & \pa_x [ \chi^{(4)}_{x} ] +\frac{1}{16} \pa_t [ Z(\rho)] +\beta_1 ,\\
   \label{chi4x}
    \chi^{(4)}_{x} & = & \frac{i}{16} \Big\{Z^{(1)}(\rho) ( q \pa_{x} \bar{q} - \bar{q} \pa_x q ) +   6 \bar{q} \pa^2_{x} q X + 4 \bar{q} \pa_{x} q \pa_x X  - 4 \pa_x ( \bar{q} \pa_{x} q) X - \pa_x ( \bar{q}  q) \pa_x X + \pa^2_x ( \bar{q} q) X + \nonumber \\
   &&  \bar{q} q \pa^2_{x}X - (\frac{1}{2} V^{(1)} +\bar{q} q )[-6 \pa_x(\bar{q} \pa^2_{x} q ) + 4 \pa^2_x (\bar{q} \pa_{x} q  ) + 4 \bar{q} \pa^3_{x} q - \pa^3_x (\bar{q} q)]  \Big\} \nonumber\\
\label{anomaly1}
\beta_1 &=&  \frac{i}{32} \( \frac{2 \bar{q} q}{V^{(1)} } + 1\)  \( \bar{q} \pa_{x}^4 q - q \pa^4_{x} \bar{q}  \) V^{(1)}
     \\     
 \label{zz1}   \frac{d}{d\rho} Z(\rho) & \equiv &   6 \int_{\rho_0}^{\rho} \hat{\rho} [ \frac{1}{2} V^{(2)}(\hat{\rho}) + 1 ] \, d\hat{\rho},
\er
where the function $\beta_1$ defines the anomaly associated with the quasi-conservation law (\ref{charge5}).  Let us write the next identity
\br
\( \bar{q} \pa_{x}^4 q - q \pa^4_{x} \bar{q}  \)  V^{(1)} &=&  \Big[ (-i \pa_t \bar{q} + \pa_x^2 \bar{q}) \pa_{x}^4 q - (i \pa_t q + \pa_x^2 q ) \pa^4_{x} \bar{q}  \Big] \\
&=& \Big[ \pa_{x} {\cal M} - i \pa_t (  \bar{q} \pa^4_x q )  \Big]     \label{id11}
    \\   
  {\cal M} &\equiv & \pa^2_x \bar{q} \pa^3_x q - \pa^3_x \bar{q} \pa^2_x q  - i \pa^3_x \bar{q} \pa_t q  + i \pa^2_x \bar{q} \pa_x \pa_t q -i \pa_x \bar{q} \pa^2_x \pa_t q + i \bar{q} \pa^3_x \pa_t q, 
\er
which is derived by using the eqs. of motion (\ref{mnls11})-(\ref{mnls22}). Therefore, using (\ref{id11}) the anomaly $\beta_1$ can be written as
\br
\label{anomaly11}
\beta_1 &=&  \frac{1}{32} \( \frac{2 \bar{q} q}{V^{(1)} } + 1\)  \Big[ \pa_{x} {\cal M} - i \pa_t (  \bar{q} \pa^4_x q )  \Big] .
\er
Next, taking into account the  relevant terms of $\chi_4$  in (\ref{chi4xt}) and the terms in  the l.h.s. of (\ref{charge5}) with partial $t-$derivatives and discarding the boundary terms with partial $x-$derivatives one can define the fifth order quasi-conserved charge 
\br
 \label{charge5a}
\frac{d}{dt} Q^{(5)}_a &=&  \int dx\, \beta_1,\\
Q^{(5)}_a & \equiv & \frac{1}{32} \int dx\,  \Big[ 2 (\bar{q} q)^3  - 8 \bar{q}q \pa_x q \pa_x \bar{q} - \bar{q}^2   (\pa_x q)^2 - q^2 (\pa_x \bar{q})^2 + \pa_x^2 \bar{q} \pa_x^2 q - 2 Z(\rho) \Big] \label{charge5a1},
\er
where the anomaly $\beta_1$ can take the form (\ref{anomaly1}) or, alternatively, the form (\ref{anomaly11}). Notice that the form of the anomaly in (\ref{anomaly1}) possesses an odd parity under (\ref{par1}) and  (\ref{aknsparity}). Therefore, one has $\int dt \int dx \, \beta_1 = 0$ implying the asymptotically conservation of the charge $Q^{(5)}_a$.   

Therefore, the fifth order eq. (\ref{charge5}) has been written  as a quasi-conservation law. Through the reduction process (\ref{nlspsi}) one can get an anomalous charge and its relevant anomaly $\beta_1$ at this order for the MNLS model, as presented in  \cite{jhep3, jhep4, jhep5}.  In fact, upon the reduction  (\ref{nlspsi})  the anomalous charge  $Q^{(5)}_a$ in (\ref{charge5a})-(\ref{charge5a1})  can be identified, dropping surface terms, to the one for the MNLS model discussed in sec. 3.6 of \cite{paper1}. 

Notice that, in the usual AKNS  limit, i.e. when $V^{(1)} = - 2 \bar{q} q$ and $V^{(2)} = - 2$ for the AKNS potential as in (\ref{nlspot}), the factor $\(\frac{2 \bar{q} q}{V^{(1)}} + 1\)$ of the anomaly $\beta_1$ in (\ref{anomaly1}) vanishes, and the term $Z(\rho)$ in the density of the charge (\ref{charge5a1}) can be set to zero (see (\ref{zz1})). Therefore, the quasi-conserved charge $Q^{(5)}_a$ in (\ref{charge5a}) becomes the fifth order charge $Q^{(5)}$  of the usual AKNS model. Actually, in this limit one has that $\chi_4 =0$ for $X=0$ (see \ref{chis}), then the r.h.s. of (\ref{charge5}) vanishes, and so, this eq. can be written as an exact conservation law.
 
So, we have constructed the set of (quasi-)conservation laws of type  (\ref{anocons}) using the Riccati-type approach  of the  modified AKNS model. By a suitable reduction process these charges can be identified to the ones of the MNLS model, as discussed above. In ref. \cite{jhep3} in the context of the anomalous Lax pair formulation of modified NLS models and through the abelianization procedure it has been constructed an infinite set of asymptotically conserved charges, which are similar in form to the exact conserved charges of the standard NLS model. 
 
\section{Dual Riccati-type formulation and novel anomalous charges}
\label{sec:dual1}

In this section we will derive the novel anomalous conservation laws through the dual formulation of the Riccati-type pseudo-potential approach. So, in order to discuss a dual formulation, let us rewrite the Riccati-type system (\ref{ricc1})-(\ref{ricc2}) and the auxiliary eq. (\ref{chi0}) as 
\br
\label{ricc1n1}
\pa_x u &=& -2 i \zeta \, u + q + \, \bar{q} \,\, u^2,\\
\label{ricc2n1}
\pa_t u &=&  2 A\, u - C\, u^2 + B - i \frac{\chi}{\bar{q}},\\
\pa_x \chi &=& \( -2 i \zeta + 2 u \bar{q} + \frac{\pa_ x \bar{q}}{\bar{q}}\) \chi + 2 \bar{q}\, u X \label{chi01},
\er
where the $A, B$ and $C$ functions are defined in (\ref{A11})-(\ref{A33}).  

Notice that the system of differential eqs. (\ref{mnls11})-(\ref{mnls22}) is invariant under the transformations: $q  \leftrightarrow \bar{q}$ and $i \leftrightarrow -i$. So, a dual formulation of the Ricati-type system (\ref{ricc1n1})-(\ref{chi01}) is achieved by performing the changes $q  \leftrightarrow \bar{q}$ and $i \leftrightarrow -i$, $u \rightarrow \bar{u}$ and $\chi \rightarrow \bar{\chi}$ into the system above. So, one has
\br
\label{ricc1d}
\pa_x \bar{u} &=& 2 i \zeta \, \bar{u} + \bar{q} + \, q \,\, \bar{u}^2,\\
\label{ricc2d}
\pa_t \bar{u} &=&  2 \bar{A}\, \bar{u} - \bar{C}\, \bar{u}^2 + \bar{B} +i \frac{\bar{\chi}}{q},\\
\label{chi0d}
\pa_x \bar{\chi} &=& \( 2 i \zeta + 2 \bar{u} q + \frac{\pa_ x q}{q}\) \bar{\chi} + 2 q\, \bar{u} X.
\er
where $\bar{u}$ is a new  Riccati-type pseudo-potential and $\bar{\chi}$ is a new  auxiliary field. Notice that the function $X$ defined in (\ref{Xanom}) remains the same. The functions $\bar{A},\bar{B}$ and $ \bar{C}$ become
\br
\label{A11d}
\bar{A} & \equiv &  2 i \zeta^2 + \frac{1}{2} i\, V^{(1)},\\
\label{A22d}
\bar{B} & \equiv & 2 \zeta \bar{q} - i \pa_x \bar{q},\\
\label{A33d}
\bar{C} & \equiv & -2 \zeta q - i \pa_x q.
\er
It is a simple calculation to verify that this dual Riccati-type system (\ref{ricc1d})-(\ref{chi0d}) reproduces the eqs. (\ref{mnls11})-(\ref{mnls22}).  

Considering  the expansions
\br
\label{expandual}
\bar{u} = \sum_{n=1}^{\infty} \bar{u}_n \, \zeta^{-n},\,\,\,\,  \bar{\chi}= \sum_{n=1}^{\infty} \bar{\chi}_n  \zeta^{-n-1},
\er
the coefficients $\bar{u}_n$ and $\bar{\chi}_n\, 's$ can be obtained from (\ref{ricc1d}) and (\ref{chi0d}), respectively.  The first components  are provided in appendix \ref{app:uchid}.

For the fields $q, \bar{q}$ and $X$ satisfying the transformation laws (\ref{aknsparity}) and (\ref{Xtr}), respectively, one can verify from the system of dual equations  (\ref{ricc1n1})-(\ref{chi01}) and (\ref{ricc1d})-(\ref{chi0d})  the following symmetry transformations 
\br
\label{ubu}
\widetilde{{\cal P}}(u) &=& - \bar{u},\,\,\,\,\,\,\,\widetilde{{\cal P}}(\bar{u}) = - u,\\
\widetilde{{\cal P}}(\chi) &=&- \bar{\chi},\,\,\,\,\,\,\, \widetilde{{\cal P}}(\bar{\chi})= - \chi.
\label{chibchi}
\er 
In fact, a careful inspection of the first six and five lowest order components for the expressions of $\{u,\bar{u}\}$  and  $\{\chi, \bar{\chi}\}$, respectively,  provided in the  appendices \ref{fsca1}, \ref{ap:chi} and \ref{app:uchid}, are in accordance, order by order in $n$, with the symmetries above, i.e.
\br
\label{ubun}
\widetilde{{\cal P}}(u_n) &=& - \bar{u}_n,\,\,\,\,\,\,\,\widetilde{{\cal P}}(\bar{u}_n) = - u_n,\,\,\,\,\,\, n=1,2,...,6;\\
\widetilde{{\cal P}}(\chi_n \pm  \bar{\chi}_n) &=& \mp  (\chi_n \pm \bar{\chi}_n),\,\,\,\,\,\,\,\,\,\,\,\,\,\,\,\,\,\,\,\,\,\,\,\,\,\,\,\,\,\,\,\,\,n=1,2,...,5;
\label{chibchin}
\er  
From the both dual systems of Riccati-type eqs. (\ref{ricc1n1})-(\ref{chi01}) and (\ref{ricc1d})-(\ref{chi0d}) one can write the next equations, respectively
\br
\label{qua1d}
\pa_t(i \bar{q} u) -\pa_x (2i \zeta \bar{q} u - \bar{q}q+ u \pa_x \bar{q}) = \chi
\er
and 
\br
\label{qua2d}
\pa_t(i q \bar{u}) -\pa_x (2i \zeta q \bar{u} + \bar{q}q - \bar{u} \pa_x q) = -\bar{\chi},
 \er
where (\ref{qua1d}) has already been considered in (\ref{qcons}) with $\chi$ defined in (\ref{qcons11}). Subtracting  the b.h.s. of (\ref{qua1d}) and (\ref{qua2d}) one has 
\br
\label{quasidual}
\pa_t [i \bar{q} u-i q \bar{u}] -\pa_x [2i \zeta (\bar{q} u -q \bar{u})- 2\bar{q}q+ u \pa_x \bar{q} + \bar{u} \pa_x q] = \chi + \bar{\chi}.
\er
Notice that the r.h.s. of the last equation is an odd expression under the special space-time operator; i.e. taking into account (\ref{chibchi}) one has $ \widetilde{{\cal P}}(\chi + \bar{\chi})= - (\chi+ \bar{\chi})$. So, the equation (\ref{quasidual}) defines  a quasi-conservation law. In fact,  the first five lowest order components  are indeed odd functions as written in (\ref{chibchin}). A usual computation shows that the components of the expansion in powers of $\zeta^{-n}$ of the quasi-conservation law (\ref{quasidual}) give rise to the normalization $(n=1)$, momentum $(n=2)$ and energy $(n=3)$ conserved charges; whereas, the higher order ones provide the same anomalous charges as the ones discussed in sec. \ref{sec:riccati}. An important observation is that the density charges of the (quasi-)conservation eq. (\ref{quasidual}) are even functions, since  the expression $[i \bar{q} u-i q \bar{u}]$ inside the partial time derivative in the l.h.s. of (\ref{quasidual}) is an even parity function.

However, the summation of the b.h.s. of (\ref{qua1d}) and (\ref{qua2d}) will not reproduce a quasi-conservation law,  since the anomaly $(\chi - \bar{\chi})$ is an even function according to  (\ref{chibchi}). In addition, in this case the expression $[i \bar{q} u+i q \bar{u}]$ of the charge density  will be an odd parity function, furnishing a trivial charge.

In the following we construct new towers of quasi-conservation laws with true anomalies, i.e. expressions  with odd parities under the symmetry transformation (\ref{par1}). Let us consider even parity expressions of the types:  $ \bar{u}u$, $i(\bar{u}\pa_x u- u \pa_x\bar{u})$,  $ \pa_x \bar{u} \pa_x u$, $ i(\bar{u}\pa_x^3 u- u \pa_x^3\bar{u})$, $i\bar{u} u (\bar{u}\pa_x u- u \pa_x\bar{u}),...$ So, one can write an infinite tower of quasi-conservation laws on top of every monomial or polynomial of these types. Next, we show the first examples of this novel set of infinite number of quasi-conservation laws.   

The next equation  follows from the both dual systems of eqs. (\ref{ricc1n1})-(\ref{chi01}) and (\ref{ricc1d})-(\ref{chi0d}) 
\br
\label{nor1}
\pa_t [\frac{1}{k}( \bar{u} u )^{k}] - \pa_x[i (\bar{u} u)^{k-1}(\bar{u} \pa_x u - u \pa_x\bar{u})] &=& {\cal A}^{(k)},\,\,\,\,\,\,\,\,\,\, k=1,2,3,...\\
{\cal A}^{(k)} &=& (\bar{u} u)^{k-1} {\cal A}-i(k-1)(\bar{u} u)^{k-2}[(\bar{u} \pa_x u)^2-(u \pa_x\bar{u})^2],\\
 {\cal A} &\equiv& 4 \zeta \bar{u} u [\bar{q} u + q \bar{u}  ] + 2 i \bar{u} u [\bar{u}\pa_x q - u  \pa_x\bar{q}  ] - 2 i \bar{u} u [\bar{q}^2u^2 - q^2 \bar{u}^2  ]+\nonumber\\
&& i [u \frac{\bar{\chi}}{q} -  \bar{u} \frac{\chi}{\bar{q}}].\label{and1}
\er
Notice that  ${\cal A}$ is an odd function, and so is the general function ${\cal A}^{(k)}$ for any positive integer $k$. Remarkably, the anomaly  ${\cal A}^{(k)}$ encompasses two types of anomalies. In fact, the last terms of ${\cal A}$ in (\ref{and1}) show the auxiliary potentials  $\bar{\chi}$ and $\chi$ which incorporate the information of the modification of the AKNS model. The remaining terms of ${\cal A}^{(k)}$ do not depend explicitly on those fields; and so, they will be present even in the standard AKNS model, i.e. for $\bar{\chi}=\chi=0$.  This is our first description, in the pseudo-potential approach, of the presence of these type of quasi-conservation laws even for a truly integrable system.         

Let us examine the first three lowest order equations  $(n=2,3,4)$, for the case $k=1$ in (\ref{nor1}).  The first two orders $n = 2, 3$ correspond, up to overall constant factors, to the exact conservation laws for the normalization and momentum charges, (\ref{n1nor}) and (\ref{n2mo}), respectively. The order $n=4$ can be written as 
\br
\nonumber
\pa_t \{\pa_x \bar{q} \pa_x q -\bar{q} \pa^2_x q - q \pa^2_x \bar{q} -2 (\bar{q} q)^2\}&-&\pa_x \{i [2(\pa_x\bar{q}\pa^2_x q-\pa_x q\pa^2_x \bar{q})-(\bar{q}\pa^3_x q- q\pa^3_x \bar{q})-\\
&& 2 \bar{q}q (\bar{q}\pa_x q - q \pa_x\bar{q})]\}= {\cal A}_0 + {\cal A}_{X}\label{n4dd}\\
{\cal A}_0 &\equiv&-2i \bar{q}q (\bar{q}\pa^2_x q - q \pa^2_x\bar{q}) -4i [(\bar{q}\pa_x q)^2 - (q \pa_x\bar{q})^2]\\
{\cal A}_X &\equiv&- 6i (\bar{q}\pa_x q - q \pa_x\bar{q}) X,
\er
with the odd functions ${\cal A}_0$ and ${\cal A}_X$ defining the relevant anomaly $({\cal A}_0+{\cal A}_X)$. Notice that ${\cal A}_X$ contains the deformation variable, $X$, and it will vanish for the standard AKNS model, i.e. ${\cal A}_X =0$. Whereas, the term ${\cal A}_0$ will be present as an anomaly even for the usual AKNS model. 
 
Next, let us examine the first two lowest order conservation laws for $k=2$ in (\ref{nor1}).  The first non-trivial quasi-conservation law is for the order $n = 4$
\br
\pa_t [(\bar{q} q)^2] - \pa_x [i \bar{q} q (\bar{q}\pa_x q - q \pa_x\bar{q})] = - i \bar{q} q [(\bar{q}\pa_x q)^2-(q \pa_x\bar{q})^2],
\er
where the r.h.s. is an odd function which does not depend explicitly on the deformation variable $X$. This quasi-conservation law corresponds, up to an overall factor $\frac{1}{4}$, to the eqs. in sec. 3.2 (with $n=2$) of the companion paper \cite{paper1} for the (modified) NLS model provided that the field identifications (\ref{nlspsi}) are performed. 

The next order $n=5$ provides
\br
\nonumber
\pa_t [i \bar{q} q (\bar{q}\pa_x q - q \pa_x\bar{q})] &-& \pa_x \{\frac{1}{2} [ \pa_x q^2 \pa_x\bar{q}^2- ((\bar{q}\pa_x q)^2-(q \pa_x\bar{q})^2)- \bar{q} q  (\bar{q}\pa^2_x q + q \pa^2_x\bar{q})-\\
\nonumber
&&\frac{4}{3} (\bar{q} q)^3]\} = [\bar{q}^2 \pa_x q \pa^2_x q + q^2 \pa_x\bar{q} \pa^2_x \bar{q} ] -[ \bar{q} (\pa_x q)^2 \pa_x \bar{q} + q \pa_x q (\pa_x \bar{q})^2] -\\
&& 2 (\bar{q} q)^2 X.
\er
In this case the anomaly contains a term with $X$ and another terms which will be present in the usual AKNS model for $X=0$.
 
In the following, let us consider the quasi-conservation law
\br
\nonumber
\pa_t [\pa_x \bar{u} \pa_x u ] - \pa_x [i (\pa_x \bar{u} \pa^2_x u - \pa_x u \pa^2_x \bar{u})] &=& 2 \zeta (\pa_x\bar{u} \pa_xq_ + \pa_x u \pa_x\bar{q}) + 2\zeta (\bar{u}^2\pa_x u \pa_x q+ u^2\pa_x \bar{u} \pa_x \bar{q})+\\ \nonumber
&&
2\zeta(\bar{q} \pa_x \bar{u} \pa_x u^2+q \pa_x u \pa_x \bar{u}^2)-
i(\pa_x \bar{q} \pa_x \bar{u} \pa_x u^2-\pa_x q \pa_x u \pa_x \bar{u}^2 )- \\ 
\label{kindual}
&&i(\pa^2_x \bar{q} \pa_x u-\pa^2_x q \pa_x \bar{u} )-i(\pa^2_x \bar{q}\, u^2\pa_x \bar{u} -\pa^2_x q\, \bar{u}^2\pa_x u) - \\
&&i (u \pa_x \bar{u}  -\bar{u} \pa_x u)\pa_x V^{(1)}+
i(\pa_x \bar{q} \pa_x \bar{u} \frac{\chi}{\bar{q}^2}-\pa_x q \pa_x u \frac{\bar{\chi}}{q^2})+\nonumber \\
&&i (\pa_x u \frac{\pa_x \bar{\chi}}{q}-\pa_x \bar{u} \frac{\pa_x \chi}{\bar{q}})-i (\pa_x \bar{u} \pa^3_x u - \pa_x u \pa^3_x \bar{u}). \nonumber
\er
The lowest order of (\ref{kindual} ) is for $n=2$ and it becomes the quasi-conservation law   
\br
\pa_t [\frac{1}{4}\pa_x \bar{q} \pa_x q ] - \pa_x [\frac{i}{4} (\pa_x \bar{q} \pa^2_x q - \pa_x q \pa^2_x \bar{q})] = \frac{i}{4} [(\bar{q} \pa_x q)^2-(q \pa_x \bar{q})^2] V^{(2)}.
\er 
Similarly, this quasi-conservation law corresponds, up to an overall factor $\frac{1}{4}$, to the eq. 3.26  (with $n=1$) in the companion paper \cite{paper1} for  the modified NLS model provided that the field identifications (\ref{nlspsi}) are performed. 

Analogous construction process for the polynomial $i(\bar{u}\pa_x u- u \pa_x\bar{u})$ will provide the momentum charge, at the lowest order conservation law ($n=2$),  and in top of it an infinite set of anomalous charges as in the tower of charges of the sec. 3.3 of the companion paper \cite{paper1}.

We have presented an unified and rigorous constructions of the first representatives of the novel type of quasi-conservation laws in the pseudo-potential approach. Let us emphasize that those quasi-conservation laws are present in the deformations of AKNS model of the type (\ref{mnls11})-(\ref{mnls22}), as well as in the standard AKNS model for potential (\ref{nlspot}). Certainly, they  deserve a more careful consideration in the context of (quasi-)integrability phenomena and, in general, the dynamics of soliton collisions, which we will address in a future work.   
 
Since the early days of integrable models the presence of an infinite number of conservation laws is among the most important features of integrability, and the presence of anomalous conservation laws, in the N-soliton sector of them, is a novelty. The anomalous charges of deformations of the SG and KdV models have also been performed in the Ricatti-type pseudo-potential approach \cite{npb1, jhep33}), reproducing the results of \cite{npb, jhep1} obtained in the zero-curvature method. In fact, this method reproduces the anomalous charges which possess the same form as the  ones of the standard integrable models. In sec. 3 of the companion  paper \cite{paper1} we have obtained infinite number of  anomalous charges of the modified NLS by a direct construction method, and in this section we have uncovered novel anomalous charges in the pseudo-potential approach. 

The partial differential equations have been regarded as infinite-dimensional manifolds and  the so-called differential coverings have been introduced, which have been used to construct some structures such as Lax pairs and B\"acklund transformations (see e.g.  \cite{krasil, igonin}). In particular, an auto-B\"acklund transformation is associated to an automorphism of the covering. Then, it would be interesting to study the properties of the dual system  (\ref{ricc1n1})-(\ref{chi01}) and (\ref{ricc1d})-(\ref{chi0d}) as some types of differential coverings of the MAKNS system  (\ref{mnls11})-(\ref{mnls22}).      

Recently, several analytical and numerical techniques have been set forward in order to study non-linear equations possessing soliton type solutions (see e.g.  \cite{jiwari} and references therein).

\section{Linear system formulation of modified AKNS system}
\label{linear}

In the next steps we will pursue a linear system of equations associated to the deformed AKNS system (\ref{mnls11})-(\ref{mnls22}). In order to tackle this problem we will resort to the Riccati-type formulation of the model presented above; so, in this context we will make use of three of the eqs. presented above,  the Riccati eq.  (\ref{ricc1}), the quasi-conservation law (\ref{qcons}) and the eq. for the auxiliary field $\chi$ in  (\ref{chi0}). Next, let us consider the transformation
\br
\label{tr12}
u = -  \frac{1}{\bar{q}} \pa_x \log{\phi}, 
\er
where $\phi$ represents a new pseudo-potential. 

With the substitution (\ref{tr12}) the Riccati eq. (\ref{ricc1}) becomes
\br
\label{r1phi}
\phi_{xx} = -[2 i \zeta  - \frac{\pa_x \bar{q}}{\bar{q}} ] \phi_x - \bar{q} q \phi .
\er

Next, consider the quasi-conservation law (\ref{qcons}) and integrate that eq. once in $x$. 
Then one gets 
\br
\label{ss}
s(x,t) = \bar{q} q - \frac{1}{ \phi}\Big[ i \phi_t - 2 i \zeta \phi_x -   \frac{\pa_x \bar{q}}{\bar{q}} \phi_x \Big],
\er
where 
\br
\label{ss1}
s(x,t) \equiv \int^x dx'  \chi.
\er

The auxiliary eq. (\ref{chi0}) upon substitution of (\ref{tr12}) becomes
\br
\label{sxx}
s_{xx} =  -2 X \pa_x \log{\phi} -  [2 i \zeta  - \frac{\pa_x \bar{q}}{\bar{q}} + 2 \pa_x \log{\phi}] s_x .  
\er
The compatibility condition $\pa_t[ \pa^2_{x} \phi ]- \pa^2_{x} [\pa_t \phi] =0$ gives rise to  the eq. of motion of the deformed AKNS (\ref{mnls11})-(\ref{mnls22}). 

Some comments are in order here. First, in the absence of deformations the auxiliary eq. (\ref{sxx}) becomes a trivial one, i.e. $\chi=X=0$ implies $s=0$. Second, one can get the linear system of eqs. (\ref{r1phi})-(\ref{ss}) as the linear formulation of the standard AKNS model, provided that $s\equiv 0$ in the l.h.s. of (\ref{ss}). 

So, following analogous constructions presented in \cite{npb1, jhep33} related to the deformations of the sine-Gordon and KdV models, we look for a linear system of eqs. associated to the deformed AKNS system.  Notice that the function $s$ in (\ref{ss})-(\ref{ss1}) will inherit from $\chi$ in (\ref{chi0}) a highly nonlinear dependence on $u$; then, through the transformation (\ref{tr12}), $s$ will have in general a nonlinear dependence on $\phi$. However, one can argue that the eq. (\ref{ss}) would represent a linear eq. for the pseudo-potential $\phi$ provided that the auxiliary field $s$ is written solely in terms of the fields $q,\bar{q}$ and $X$ and their derivatives. So, let us assume the next Ansatz
\br
\label{lin01}
\pa_x \phi &=& {\cal A}_x \phi,\\
\label{lin02}
\pa_t \phi &=& {\cal A}_t \phi,
\er
where the functions ${\cal A}_x$ and ${\cal A}_t$ represent the gauge connections and depend on the fields of the model. The compatibility condition of this  system will provide the eq. of motion
\br
\label{eqm}
\pa_{t} {\cal A}_x - \pa_x {\cal A}_t =0.
\er  
Using (\ref{lin01}) into (\ref{r1phi}) one gets the following Riccati eq. for ${\cal A}_x$
\br
\label{riccatiAx}
\pa_x {\cal A}_x =  - 2i \zeta  {\cal A}_x - ({\cal A}_x)^2 - \bar{q} q + \frac{\pa_ x \bar{q}}{\bar{q}} {\cal A}_x .
\er
Likewise, replacing (\ref{lin01})-(\ref{lin02}) into  (\ref{ss}) one gets a relationship for the quantity $s$
\br
\label{ss110}
s = \bar{q} q - i {\cal A}_t + ( 2 i  \zeta  +\frac{\pa_ x \bar{q}}{\bar{q}}) {\cal A}_x .
\er
Substituting this form of $s$ into the eq. (\ref{sxx}) and using the eqs.  (\ref{eqm})-(\ref{riccatiAx})  one gets the eq. of motion of the deformed AKNS (\ref{mnls11})-(\ref{mnls22}). So, the form of $s$ in (\ref{ss110}) is consistent with the dynamics of the deformed model.

Notice that the system of eqs.  (\ref{lin01})-(\ref{lin02}) are defined up to a gauge transformation of the type
\br
\label{gauge1}
\phi &\rightarrow& e^{\Lambda} \phi\\
\label{gauge2}
{\cal A}_x &\rightarrow&  {\cal A}_x + \pa_x \Lambda\\
\label{gauge3}
{\cal A}_t &\rightarrow& {\cal A}_t + \pa_t \Lambda,
\er
for an arbitrary function $\Lambda$. We will discuss this point below in more detail.

In order to find a linear system formulation of the modified AKNS it is needed a certain amount of guesswork out of the eq. (\ref{ss}). Moreover, due to the gauge symmetry (\ref{gauge1})-(\ref{gauge3}) it is possible to make a particular choice for the connections ${\cal A}_x$ and ${\cal A}_t$. Let us propose the following linear system of equations as the linear formulation of the deformed AKNS \footnote{Below we will provide a gauge transformation between the systems (\ref{sys1})-(\ref{sys2}) and (\ref{lin01})-(\ref{lin02}).}
\br
\label{sys1}
\pa_t \Phi &=& A_t \Phi\\
\label{sys2}
\pa_x \Phi &=& A_x \Phi\\
\label{AA1}
A_x &\equiv & -\zeta \pa_x\bar{q} + 2 \bar{q} \,\(\frac{a_0 \zeta +\zeta^2 a_1 }{2 \zeta \bar{q}+ i \pa_x\bar{q}}\),\\
A_t &\equiv & \frac{b_0 +\zeta  b_1}{2 i \zeta^2\bar{q} - \zeta \pa_x\bar{q}} + \zeta \, \int^{x}\, dx' \, \frac{b_0 + b_1 \zeta + b_2 \zeta^2}{(2 \zeta \bar{q}+ i \pa_{x'}\bar{q})^2},\label{AA2}
\er
such that 
\br
\nonumber
b_0 &\equiv& \frac{1}{2} i \pa_x \bar{q}^2 (2 \pa_t a_0 + \pa_x V^{(1)} \pa_x \bar{q} ) +i(\pa_x \bar{q})^2 (V^{(1)} \pa_x\bar{q} -\pa^3_x \bar{q})+2 a_0 (\bar{q}^2 \pa_x V^{(1)} + \pa_x \bar{q} \pa^2_x\bar{q} - \bar{q} \pa^3_x \bar{q})\\
b_1 &\equiv& 2 \bar{q}^2 [2 \pa_t a_0 + \pa_x V^{(1)} (a_1 + 2 \pa_x \bar{q})] + 2 a_1 \pa_x \bar{q} \pa^2_x \bar{q} + \bar{q} [2i \pa_t a_1 \pa_x \bar{q} + 4 V^{(1)} (\pa_x \bar{q})^2-2 (a_1 + 2 \pa_x \bar{q} )\pa^3_x\bar{q}],\nonumber\\
b_2 &\equiv& 8 \bar{q}^2 [\pa_t a_1 - i (\bar{q} \pa_x V^{(1)} + V^{(1)} \pa_x \bar{q} - \pa^3_x\bar{q})],\nonu
\er  
where $a_0$ and $a_1$ are some nonvanishing auxiliary functions.
The compatibility condition of the system of eqs. (\ref{sys1})-(\ref{sys2}); i.e. $\pa_t \pa_x (\Phi)-\pa_x \pa_t (\Phi)=0$, furnishes the next equation
\br
\label{eqm1}
\pa_{t} A_x - \pa_x A_t =0.
\er
Substituting the expressions of the connection components  $\{A_x\,,\, A_t\}$ defined in (\ref{AA1})-(\ref{AA2}) into (\ref{eqm1})  one gets the next expression, which is a polynomial in powers of $\zeta$ 
\br
\label{pol1}
\{ 8 i \bar{q}^2\pa_x(V^{(1)} \bar{q} + i \pa_t \bar{q} - \pa^2_x \bar{q})\} \zeta^2 -\\ 
\label{pol2}
 \{2 a_1 [\pa_x \bar{q}(\pa^2_x \bar{q}-i \pa_t \bar{q}) + \bar{q} (\bar{q} \pa_x V^{(1)} + i \pa_x \pa_t \bar{q} - \pa^3_x \bar{q}) ] +2 \pa_x\bar{q}^2 \pa_x[V^{(1)} \bar{q} + i \pa_t \bar{q} - \pa^2_x \bar{q}   ]   \} \zeta-\\
\{2 a_0 [\pa_x  \bar{q} (\pa^2_x \bar{q}  - i \pa_t \bar{q} ) + \bar{q}  (\bar{q}  \pa_x V^{(1)} + i \pa_x \pa_t \bar{q}  - \pa^3_x \bar{q} )] + i (\pa_x \bar{q} )^2\pa_x (V^{(1)} \bar{q} + i \pa_t \bar{q} - \pa^2_x \bar{q})\} \equiv 0
\label{pol3}
\er 
Therefore, equating to zero the coefficient of $\zeta^2$ in (\ref{pol1}) provides the identity
\br
\pa_x[ V^{(1)} \bar{q} + i \pa_t \bar{q} - \pa^2_x \bar{q}] = 0.
\er
Next, replacing this identity into the coefficients of $\zeta$ and $\zeta^{0}$ in the eqs. (\ref{pol2}) and (\ref{pol3}), respectively,  one can get 
\br
\label{dakns21}
a_i\, [-i \pa_t \bar{q} + \pa^2_x \bar{q} - V^{(1)} \bar{q}] =0,\,\,\,\,\,i=0,1.
\er
Since the auxiliary fields $a_i$ are non-vanishing arbitrary functions one gets the second eq. (\ref{mnls22}) of the AKNS system. The first eq. (\ref{mnls11}) we will derive below.
 
Next, let us consider the linear system \footnote{Notice that the connections from (\ref{sys11})-(\ref{sys21}) and (\ref{sys1})-(\ref{sys2}) can be related as $\widetilde{{\cal A}}_{\mu} = \widetilde{{\cal P}} ( {\cal A}_{\mu} ) \,\,(\mu = \{x, t\}),\,\,\,\widetilde{a}_i =  \widetilde{{\cal P}}(a_i)\,\, (i=0,1)$, where $\widetilde{{\cal P}} $ is the parity transformation defined  in (\ref{par1}) and (\ref{aknsparity}).}
\br
\label{sys11}
\pa_t \widetilde{\Phi} &=& \widetilde{A}_t \widetilde{\Phi}\\
\label{sys21}
\pa_x \widetilde{\Phi} &=& \widetilde{A}_x \widetilde{\Phi}\\
\label{cA1}
\widetilde{A}_x &\equiv &  \zeta \pa_x q + 2 q \,\(\frac{\widetilde{a}_0 \zeta +\zeta^2 \widetilde{a}_1 }{2 \zeta q- i \pa_x q}\),\\
\label{cA2}
\widetilde{A}_t &\equiv & \frac{\bar{b}_0 +\zeta  \bar{b}_1}{-2 i \zeta^2 q - \zeta \pa_x q}  -\zeta \, \int^{x}\, dx' \, \frac{\widetilde{b}_0 + \widetilde{b}_1 \zeta + \widetilde{b}_2 \zeta^2}{(2 \zeta q - i \pa_{x'}q)^2},
\er
with
\br
\nonumber
\widetilde{b}_0 &\equiv& \frac{1}{2} i \pa_x q^2 (2 \pa_t \widetilde{a}_0 - \pa_x V^{(1)} \pa_x q ) - i(\pa_x q)^2 (V^{(1)} \pa_x q -\pa^3_x q)-2 \widetilde{a}_0 (q^2 \pa_x V^{(1)} + \pa_x q \pa^2_x q - q \pa^3_x q)\\
\widetilde{b}_1 &\equiv& -2 q^2 [2 \pa_t \widetilde{a}_0 + \pa_x V^{(1)} (\widetilde{a}_1 - 2 \pa_x q)] - 2 \widetilde{a}_1 \pa_x q \pa^2_x q + q [2i \pa_t \widetilde{a}_1 \pa_x q + 4 V^{(1)} (\pa_x q)^2+2 (\widetilde{a}_1 - 2 \pa_x q )\pa^3_x q],\nonumber\\
\widetilde{b}_2 &\equiv& -8 q^2 [\pa_t \widetilde{a}_1 - i (q \pa_x V^{(1)} + V^{(1)} \pa_x q - \pa^3_x q)].\nonu
\er  
	 
The compatibility condition of the system of eqs. (\ref{sys11})-(\ref{sys21}); i.e. $\pa_t \pa_x (\widetilde{\Phi})-\pa_x \pa_t (\widetilde{\Phi})=0$, furnishes the next polynomial in powers of $\zeta$ 
\br
\label{pol11}
\{ 8 i q^2\pa_x(V^{(1)} q + i \pa_t q - \pa^2_x q)\} \zeta^2 -\\ 
\label{pol21}
 \{2 \widetilde{a}_1 [\pa_x q(\pa^2_x q-i \pa_t q) + \bar{q} (q \pa_x V^{(1)} + i \pa_x \pa_t q - \pa^3_x q) ] +2 \pa_x q^2 \pa_x[V^{(1)} q+ i \pa_t q - \pa^2_xq   ]   \} \zeta-\\
\{2 \widetilde{a}_0 [\pa_x  q (\pa^2_x q - i \pa_t q ) + q  (q  \pa_x V^{(1)} + i \pa_x \pa_t q - \pa^3_x q )] + i (\pa_x q )^2\pa_x (V^{(1)} q + i \pa_t q - \pa^2_x q)\} \equiv 0.
\label{pol31}
\er 
Therefore, equating to zero the coefficient of $\zeta^2$ in (\ref{pol11}) provides the identity
\br
\pa_x [V^{(1)}q - i \pa_t q- \pa^2_x q] = 0.
\er
Next, replacing this identity into the coefficients of $\zeta$ and $\zeta^{0}$ in the eqs. (\ref{pol21}) and (\ref{pol31}), respectively,  one can get 
\br
\label{dakns2}
\widetilde{a}_i\, [i \pa_t q + \pa^2_x q - V^{(1)} q]  =0,\,\,\,\,\,i=0,1.
\er
Since the auxiliary fields $\widetilde{a}_i$ are nonvanishing arbitrary functions one gets the first eq. (\ref{mnls11}) of the AKNS system.

Therefore, the linear formulation (\ref{sys11})-(\ref{sys21}) is related to the first deformed AKNS eq.  (\ref{mnls11}), whereas the linear formulation  (\ref{sys1})-(\ref{sys2})  is related to the second deformed AKNS eq. (\ref{mnls22}). These separate formulations can be joined together  into just one linear system defined as
\br
\label{ss11}
\pa_x\(\begin{array}c
\Phi \\
\widetilde{\Phi} \end{array}\) &=& {\cal M} \, \(\begin{array}c
\Phi \\
\widetilde{\Phi} \end{array}\),\,\,\,\,\,\, {\cal M} \equiv \(\begin{array}{cc}
A_x  & 0\\
0  & \widetilde{A}_x \end{array}\) \\
\pa_t\(\begin{array}c
\Phi \\
\widetilde{\Phi} \end{array}\) &=& {\cal N}\, \(\begin{array}c
\Phi \\
\widetilde{\Phi} \end{array}\) ,\,\,\,\,\,\,{\cal N} \equiv  \(\begin{array}{cc}
A_t  & 0\\
0  & \widetilde{A}_t \end{array}\)\,.\label{ss22}
\er
So,  the compatibility condition of this system provides the zero-curvature eq.
\br
\label{zc1}
\pa_{t} {\cal M}  - \pa_{x} {\cal N} + \Big[{\cal M} ,\,  {\cal N} \Big] = 0. 
\er
Notice that  ${\cal M}$ and ${\cal N}$ are diagonal matrices, and so, one has $[{\cal M} ,\,  {\cal N} ]=0$; therefore, the linear formulation (\ref{ss11})-(\ref{ss22}) splits into the relevant formulations in (\ref{sys1})-(\ref{sys2}) and  (\ref{sys11})-(\ref{sys21}), respectively. Then, the eqs. of motion of the deformed AKNS model (\ref{mnls11})-(\ref{mnls22}) can be obtained from (\ref{zc1}).  

For completeness we provide a gauge transformation between the system (\ref{lin01})-(\ref{lin02}) and the  system  (\ref{sys1})-(\ref{sys2}). So, the gauge transformation  (\ref{gauge1})-(\ref{gauge3}) can be written as
\br
\phi &=& e^{-\Lambda} \Phi\\
A_x &=& {\cal A}_x + \pa_x \Lambda\\
A_t &=& {\cal A}_t + \pa_t \Lambda,
\er
where $\Omega \equiv \pa_x\Lambda$ satisfies the Riccati eq. 
\br
\pa_x \Omega =2 \Omega^2 -  (2i \zeta + 2 A_x - \frac{\pa_x \bar{q}}{\bar{q}} )\,  \Omega + i \zeta A_x + \frac{1}{2} A^2_x + \frac{1}{2} \bar{q}q + \frac{1}{2} \pa_x A_x - \frac{1}{2} A_x \frac{\pa_x \bar{q}}{\bar{q}}.
\er 
A similar construction can be performed for the gauge transformation of the sector with the connection $\(\widetilde{A}_x,\, \widetilde{A}_t\)$.  In fact,  in order to perform  that gauge transformation for  the full system  (\ref{ss11})-(\ref{ss22}) we can have
\br
\(\begin{array}c
\phi \\
\widetilde{\phi} \end{array}\) &=&g \,\(\begin{array}c
\Phi \\
\widetilde{\Phi} \end{array}\),\,\,\,\,\,\,g \equiv  \(\begin{array}{cc}
  e^{-\Lambda} & 0\\
0  & e^{-\widetilde{\Lambda}} \end{array}\) \\
A_{\mu} &=&  {\cal A}_{\mu} +  \pa_{\mu} \Lambda\\
\widetilde{A}_{\mu} &=& \widetilde{{\cal A}}_{\mu} + \pa_{\mu} \widetilde{\Lambda},\,\,\,\,\mu = \{x, t\}.
\er
Since the matrices ${\cal M} $\, , $ {\cal N}$\, and $g$ are diagonal  the connection components  from the both sectors do not couple in the process.

Actually, all of the constructions above can be reproduced for the modified NLS model (\ref{nlsd}) since one can perform the reduction process of the modified AKNS model ($ MAKNS \rightarrow MNLS$)  through the identifications (\ref{nlspsi}). 

\subsection{Infinite set of non-local conserved charges}
\label{sec:nonlocal}

For each of the linear systems in (\ref{sys1})-(\ref{sys2}) and (\ref{sys11})-(\ref{sys21}) it is possible to construct a set of non-local conserved charges. The construction of analogous  linear systems and their associated non-local charges have recently been performed for some deformations of the sine-Gordon and KdV models \cite{npb1, jhep33}. In fact, following an iterative technique developed by  Br\'ezin et.al. \cite{brezin}, the authors in \cite{npb1, jhep33} have uncovered infinite set of non-local conservation laws for the relevant linear systems associated to the  deformations of the SG and KdV models, respectively. So, in order to make this paper self-contained we summarize the main points of the construction since the procedure is quite similar to the ones undertaken for the SG and KdV models.  So, let us define the currents 
\br
J_{\mu}^{(n)} &=& \frac{\pa}{\pa x_\mu} \chi^{(n)},\,\,\,x_\mu \equiv x, t;\,\,\,\,n=0,1,2,...\\
d \chi^{(1)} &=& A_{\mu} dx_{\mu}\\
&\equiv& A_x dx + A_t dt ,\\
J_{\mu}^{(n+1)} &=& \frac{\pa}{\pa x_\mu} \chi^{(n)}-A_{\mu} \chi^{(n)};\,\,\,\,\,\chi^{(0)}=1.
\er
Next, an inductive procedure is used to show that the  (non-local) currents $J_{\mu}^{(n)}$ are conserved
\br
\label{nlcl}
\pa_{t} J^{(n)}_{t} - \pa_{x} J^{(n)}_{x} =0,\,\,\,\,n=1,2,3,...
\er
In fact, the first non-trivial  current becomes $J_{\mu}^{(1)}=(A_x, A_t)$ whose conservation law $\pa_t A - \pa_x A_t=0$  reproduces  the eq.  (\ref{eqm1}). The second  order current becomes $J_{\mu}^{(2)}=(A_x - A_x\chi^{(1)},A_t-A_t \chi^{(1)})$, and from the conservation law (\ref{nlcl}), using the first order conservation law (\ref{eqm1}), one gets
\br
\label{j2}
\pa_t [A_x \chi^{(1)}] - \pa_x [A_t\chi^{(1)}]=0.
\er
The third order current becomes $J_{\mu}^{(3)}=(\frac{\pa}{\pa x}\chi^{(2)}-A_x\chi^{(2)}, \frac{\pa}{\pa t}\chi^{(2)}-A_t\chi^{(2)})$. So, at this order, the conservation law  (\ref{nlcl}), upon using the first (\ref{eqm1}) and second (\ref{j2}) order conservation laws, can be written as 
\br
\label{j3}
\pa_t [A_x \chi^{(2)}] - \pa_x [A_t \chi^{(2)}]=0.
\er  
where
\br
\pa_ x \chi^{(2)}= A_x - A_x\chi^{(1)},\,\,\,\,\pa_ t \chi^{(2)}= A_t-A_t\chi^{(1)}.
\er
Then, one can write the infinite tower of non-local conservation laws as 
\br 
\label{nonl1}
\pa_t [ A_x \chi^{(1)} ] - \pa_x [ A_t \chi^{(1)} ] &=& 0,\\
\label{nonl2}
\pa_t [ A_x \chi^{(n)} ] - \pa_x [ A_t\chi^{(n)} ] &=&0,\,\,\,\,\,n=2,3,4,...\\
\label{nonl3}
\pa_x \chi^{(n)} &=& A_x - A_x \chi^{(n-1)},\,\,\,\,\,\,\pa_t \chi^{(n)} = A_t - A_t\chi^{(n-1)}.
\er 
A similar procedure can be performed for the sector with the gauge connection $\(\widetilde{A}_x,\, \widetilde{A}_t\)$ in (\ref{cA1})-(\ref{cA2}), giving rise to another tower of infinite number  of non-local conservation laws
\br 
\label{nonl11}
\pa_t [ \widetilde{A}_x \widetilde{\chi}^{(1)} ] - \pa_x [ \widetilde{A}_t \widetilde{\chi}^{(1)} ] &=& 0,\\
\label{nonl21}
\pa_t [ \widetilde{A}_x \widetilde{\chi}^{(n)} ] - \pa_x [ \widetilde{A}_t\widetilde{\chi}^{(n)} ] &=&0,\,\,\,\,\,n=2,3,4,...\\
\label{nonl31}
\pa_x \widetilde{\chi}^{(n)} &=&\widetilde{A}_x - \widetilde{A}_x \widetilde{\chi}^{(n-1)},\,\,\,\,\,\,\pa_t \widetilde{\chi}^{(n)} = \widetilde{A}_t - \widetilde{A}_t\widetilde{\chi}^{(n-1)}.
\er 
Due to the reduction process $ MAKNS \rightarrow MNLS$,  through the identification (\ref{nlspsi}),  the towers of non-local charges constructed above can directly be reproduced for the modified NLS model (\ref{nlsd}).  Moreover, additional reductions of the standard AKNS system have been reported which define some integrable non-local NLS, SG  and KdV  models \cite{ablowitz, reverse}. So, in the context of the modified AKNS the relevant NLS-type, SG-type and  KdV-type equations will appear for a convenient choice of the parameters $A, B$ and $C$, as well as the auxiliary fields like $r$ and $s$, in the MAKNS system  (\ref{ricc1})-(\ref{ricc2}). The suitable choices have been done in \cite{npb1} for the modified SG-like and in \cite{jhep33} for the modified KdV-like systems, respectively. So, our calculations and results above can be reproduced for the following reductions of the MAKNS system (\ref{ricc1})-(\ref{ricc2})
\br
\label{mnls44}
\bar{q}(x,t) & \equiv &\sigma \, q^{\star}(x, t), \,\,\,\,\,\,\,\,\, \sigma = \pm 1,\,\,\,\,\,\star \equiv  \mbox{complex conjugation}\\
\label{tmnls}
\bar{q}(x,t) & \equiv &\sigma \, q(x, -t), \\
\label{xmnls}
\bar{q}(x,t) & \equiv &\sigma \, q^{\star}(-x, t), \\
\label{xtmnls}
\bar{q}(x,t) & \equiv &\sigma \, q(-x, -t),  \,\,\,\,q\in \IC\\
\label{xtcmnls}
\bar{q}(x,t) & \equiv &\sigma \, q^{\star}(-x, -t),  \\
\label{xtrmnls}
\bar{q}(x,t) & \equiv &\sigma \, q(-x, -t),\,\,\,\,q\in \IR.
\er
The first reduction (\ref{mnls44}) is just the reduction $ MAKNS \rightarrow MNLS$  (\ref{nlspsi}) we have discussed in this paper. 
We expect that the second, third and fourth reductions above will give rise to non-local MNLS-type equations and the
last two of them to non-local modified KdV-type evolution equations. The construction of the anomalous charges and the symmetries satisfied by the relevant anomalies associated to the  NLS-type (\ref{tmnls})-(\ref{xtmnls}) and (real or complex) KdV-type (\ref{xtcmnls})-(\ref{xtrmnls}) reductions deserve careful analysis and we will postpone those important issues for future research. 

So far, the relevant deformations of the SG and KdV models have been considered in the literature \cite{npb1, jhep33}, and  they share a similar structure regarding their non-local conservation laws with the one of the AKNS-type models in (\ref{nonl1})-(\ref{nonl3}), since their linear formulations possess the same form as in (\ref{sys1})-(\ref{sys2}) or (\ref{sys11})-(\ref{sys21}). Since the algebra of conserved charges in certain two-dimensional integrable quantum field theories is also present in the classical theory as a Poisson-Hopf algebra \cite{mackay1, mackay2}, it would be interesting to search for those type of classical Yangian algebras related to the set of non-local currents and charges for the deformations of the integrable models. The non-local conserved charges, as in the non-linear $\sigma-$model, would be relevant at the quantum level and they would imply absence of particle production (see e.g. \cite{abdalla, luscher}).

The same technique as above can be used to construct infinite sets of non-local conserved charges for deformations of other integrable models. The issues such as  the independence and  involution of the charges deserve further analysis and remain as open questions. In order to examine those properties one must search for a Hamiltonian formulation of the model and compute the relevant Poisson-like brackets among the charges. An interesting question arises about the algebra of these infinite number of nonlocal charges. One expects the algebra to be nonlinear very much like the ones of  the nonlinear sigma model and topological field theory (see e.g. \cite{das} and references therein).    

The AKNS-type models are quite ubiquitous in the nonlinear science and it would be interesting to investigate the relevance and physical consequences of the infinite towers of infinitely many anomalous and non-local charges discussed in this paper. Some remarkable and profound relationships  between integrable models and gauge theories have been uncovered in recent years. For example, it has been proposed a kind of triality among gauge theories, integrable models and gravity theories in some UV regime. In particular,  the $(1+1)D$ nonlinear Schr\"odinger equation corresponds to the $2D\, {\cal N} = (2, 2)^{\star} U(N )$super  Yang-Mills theory (see \cite{nian} and references therein).   

We have used the Mathematica software for the various symbolic computations. The symbolic computations turn out to be usefull in non-linear physics, for example to get soliton solutions (see e.g. \cite{ref5}).
    
\section{Conclusions and discussion}
 \label{ap:conclu}
  
We have performed the Riccati-type pseudo-potential approach to deformations of the AKNS model in sec. \ref{sec:riccati}, such that the modified NLS is obtained through a certain reduction. In this framework it has been  constructed infinite towers of quasi-conservation laws and discussed their properties and relationships with the MNLS model. This construction reproduced the tower of NLS-type quasi-conserved charges obtained  in the anomalous zero-curvature approach of \cite{jhep3}.  Moreover, in sec. \ref{sec:dual1} we have introduced a dual Riccati-type pseudo-potential approach and uncovered, in that framework, a novel set of infinite number of quasi-conservation laws, such that it encompasses the quasi-conservation laws obtained by a direct method starting from the eqs. of motion in sec. 3 of the companion paper \cite{paper1}. 

In the framework of the Riccati-type pseudo-potential approach we have constructed a couple of  linear systems of equations, (\ref{sys1})-(\ref{sys2}) and  (\ref{sys11})-(\ref{sys21}), whose relevant compatibility conditions give rise to the modified AKNS system of equations  (\ref{mnls11})-(\ref{mnls22}). The second system of linear eqs. (\ref{sys11})-(\ref{sys21}) is related to the first one  (\ref{sys1})-(\ref{sys2})  through the transformation  (\ref{par1}) and (\ref{aknsparity}). In subsection \ref{sec:nonlocal} we have constructed two towers of  infinite sets of non-local conservation laws associated to the linear formulations, respectively. These linear systems and their associated non-local charges deserve more careful considerations; in particular, regarding their relationships of their associated non-local currents with  the so-called classical Yangians \cite{mackay1, mackay2}. 

In view of the current results, on deformations of SG, KdV and in this paper on deformations of the AKNS and related NLS models, one can inquire about the non-local properties of the quasi-integrable systems studied in the literature, such as the deformations of the Bullough-Dodd, Toda and SUSY sine-Gordon systems \cite{jhep6, toda, epl}, and more specific structures, such as the complete list of the towers of infinite number of anomalous charges and the (non-local) exact conservation laws, as discussed in this paper. Moreover, it would be interesting to consider the general AKNS model and study their (non-local) reductions (\ref{tmnls})-(\ref{xtrmnls}) as proposed  in  \cite{reverse, ablowitz}, as well as their relevant deformations in the lines discussed above.  
  
\section{Acknowledgments}

HB thanks FC-UNI (Lima-Per\'u) and FC-UNASAM (Huaraz-Per\'u) for hospitality during the initial stage of the work. MC thanks the Peruvian agency Concytec for partial financial support. LFdS thanks CEFET Celso Sukow da Fonseca-Rio de Janeiro-Brazil for kind support. The authors thank A. C. R. do Bonfim, H. F. Callisaya, C. A. Aguirre, J. P. R. Campos, R. Q. Bellido, J.M.J. Monsalve and A. Vilela for useful discussions.  
 
\appendix

\section{The $u_n'$s of the first set of charges}
\label{fsca1}

The $u_n'$s can be determined recursively by substituting the expansion (\ref{expan}) into (\ref{ricc1}). Then the first quantities become
\br
u_1 &=& -\frac{1}{2} i q\\
u_2 &=& \frac{1}{2} i \pa_x u_1\\
u_3 &=& \frac{1}{2} i \( - \bar{q} u_1^2 + \pa_x u_2\)\\
u_4 &=& \frac{1}{2} i \( - 2 \bar{q} u_1 u_2 + \pa_x u_3\)\\
u_5 &=& \frac{1}{2} i \( -  \bar{q} u_2^2 - 2 \bar{q} u_1 u_3 + \pa_x u_4\)\\
u_6 &=& \frac{1}{2} i \( -  2 \bar{q} (u_3 u_2 + u_1 u_4) + \pa_x u_5\)\\
&&............................\nonumber
\er
The above sequence can be written for any $n$ (even or odd) as follows
\br
u_n & = & \frac{1}{2} i \Big[ -2 \bar{q} \sum_{ \begin{array}c i_1+i_2 = n-1\\
i_1 \neq i_2 \end{array}}  u_{i_1}  u_{i_2}  +\pa_x u_{n-1} \Big],\,\,\,\,\,\,\,\,\,\,\,\,\,\,\,\,\,\,\,\,\,\,\,\,\,\,\,\,\,\,\,\,n = even\\
u_n & = & \frac{1}{2} i \Big[ - \bar{q} \,  u_{\frac{n-1}{2}}^2 -2 \bar{q} \sum_{\begin{array}c i_1+i_2 = n-1\\
i_1 \neq i_2 \end{array} } \, u_{i_1}  u_{i_2}  +\pa_x u_{n-1} \Big],\,\,\,\,\,\,n = odd.
\er  
In terms of the field components the first six $u_i\,(i=1,2,...6)$ become
\br
u_1 &=& -\frac{1}{2} i q\\
u_2 &=& \frac{1}{4}  \pa_x q\\
u_3 &=&\frac{1}{8} i \( \bar{q} q q + \pa^2_x q \)\\
u_4 &=&-\frac{1}{16}  [ 4 \bar{q} q \pa_x q + q q \pa_x \bar{q}  + \pa^3_x q ]\\
u_5 &=&-\frac{1}{32} i [ 2 (\bar{q} q)^2 q + 5 \bar{q} (\pa_x q)^2 + 6 q \( \pa_x q \pa_x \bar{q} + \bar{q}  \pa^2_x q\) + q^2 \pa^2_x \bar{q} + \pa^4_x q ]\\
u_6 &=& \frac{1}{64} \Big[6 q^3  \bar{q} \pa_x \bar{q}  +11 \pa_x \bar{q} (\pa_x q)^2  + 18  \bar{q} \pa_x q \pa^2_x q + 4 q ( 3 \pa^2_x q  \pa_x \bar{q} + 2 \pa_x q \pa^2_x \bar{q} +2 \bar{q} \pa^3_x q ) + \nonumber \\
&& q^2 (16 \bar{q} ^2 \pa_x q +  \pa^3_x \bar{q} ) + \pa^5_x q  \Big].
\er

\section{The $\chi$ components}

\label{ap:chi}
The components of the expansion of $\chi$ in a recursive form become
\br
\chi_1 &=& - i \bar{q}  u_1 X\\
\chi_2 &=& -i  \bar{q}  u_2 X  + \frac{1}{2} i \pa_x \chi_1 -   \frac{1}{2} i \chi_1 \frac{\pa_x \bar{q} }{\bar{q}} \\
\chi_3 &=& -i  \bar{q}  u_3 X  -  i  \bar{q}  u_1 \chi_1  + \frac{1}{2} i \pa_x \chi_2 -   \frac{1}{2} i \chi_2 \frac{\pa_x \bar{q} }{\bar{q}} \\
\chi_4 &=& -i  \bar{q}  u_4 X    -i  \bar{q}  u_2 \chi_1  -i  \bar{q}  u_1 \chi_2 + \frac{1}{2} i \pa_x \chi_3 -   \frac{1}{2} i \chi_3 \frac{\pa_x \bar{q} }{\bar{q}} \\
\chi_5 &=& -i  \bar{q}  u_5 X    -i  \bar{q}  u_3 \chi_1   - i  \bar{q}  u_2 \chi_2 - i  \bar{q}  u_1 \chi_3 + \frac{1}{2} i \pa_x \chi_4 -   \frac{1}{2} i \chi_4 \frac{\pa_x \bar{q} }{\bar{q}} \\
&&.............\nonumber
\er
The above sequence can be written for any $n$ as
\br
\chi_n &=& - i\,  \bar{q}  u_n X    -i \bar{q}  \sum_{i_1+i_2 = n-1} u_{i_1} \chi_{i_2}  + \frac{1}{2} i \,\pa_x \chi_{n-1} -   \frac{1}{2} i \chi_{n-1} \frac{\pa_x \bar{q} }{\bar{q}};\,\,\,\,\,\,\,\,\,  \chi_0\equiv 0,\,\,\,\,\,\,\,n=1,2,....
\er
The first five components become
\br  \nonumber
\chi_1 &=& -\frac{1}{2}  \bar{q} q X.\\ \nonumber
\chi_2 &=& -\frac{1}{4} i [2 \bar{q}  \pa_x q X +  \bar{q}  q \pa_x X].\\
\label{chis}
\chi_3 &=& \frac{1}{8}  [3 \bar{q}  \pa_x q \pa_x X + 3 \bar{q}  X ( \bar{q}  q^2 + \pa^2_x q) +  \bar{q}  q \pa^2_x X].\\
\nonumber
\chi_4 &=&\frac{i}{16} [  \bar{q} q^2\( 5 \bar{q} \pa_x X + 4 X \pa_x  \bar{q} \) + 6   \bar{q}  \pa^2_x q \pa_x X + 4 \bar{q} \pa_x q \pa^2_x X + 4 \bar{q} X \pa^3_x q + \bar{q} q \( 16 X \bar{q} \pa_x q + \pa^3_x X \) ]. 
\\ 
\nonumber
\chi_5 &=& -\frac{1}{32} \bar{q} \Big\{ 10  q^3 \bar{q}^2 X + q^2 \( 9 \pa_x X \pa_x  \bar{q} + 7 \bar{q} \pa^2_x X + 5 X \pa_x^2 \bar{q} \) + \\
\nonumber
&& 5  [ 2 \pa^2_x q \pa^2_x X + 2 \pa^3_x q \pa_x X +  \pa_x^3 X \pa_x q + X (5 \bar{q} (\pa_x q)^2 + \pa^4_x q)]+ \\
\nonumber
&& q [ 2 \pa_x q (17 \bar{q} \pa_x X + 15 X \pa_x  \bar{q} ) + 30 X  \bar{q} \pa^2_x q + \pa^4_x X ] \Big\}. 
\er

\section{The first $\bar{u}_n$ and $\bar{\chi}_n$}
\label{app:uchid}

Next we provide the lowest order components of the dual potentials $\bar{u}$ and $\bar{\chi}$
\br
\bar{u}_1 &=& \frac{1}{2} i \bar{q},\\
\bar{u}_2 &=& \frac{1}{4}  \pa_x \bar{q},\\
\bar{u}_3 &=&-\frac{1}{8} i \( q \bar{q}\bar{q} + \pa^2_x \bar{q} \),\\
\bar{u}_4 &=&-\frac{1}{16}  [ 4 q \bar{q} \pa_x \bar{q} + \bar{q} \bar{q} \pa_x q  + \pa^3_x \bar{q} ],\\
\bar{u}_5 &=&\frac{1}{32} i [ 2 (\bar{q}q )^2 \bar{q} + 5 q (\pa_x \bar{q})^2 + 6 \bar{q} \( \pa_x \bar{q} \pa_x q + q  \pa^2_x \bar{q}\) + \bar{q}^2 \pa^2_x q + \pa^4_x \bar{q} ] ,\\
\bar{u}_6 &=& \frac{1}{64} \Big[6 \bar{q}^3  q \pa_x q  +11 \pa_x q (\pa_x \bar{q})^2  + 18  q \pa_x \bar{q} \pa^2_x \bar{q} + 4 \bar{q} ( 3 \pa^2_x \bar{q}  \pa_x q + 2 \pa_x \bar{q} \pa^2_x q +2 q \pa^3_x \bar{q}) + \nonumber \\
&& \bar{q}^2 (16 q^2 \pa_x \bar{q} +  \pa^3_x q) + \pa^5_x \bar{q}  \Big]
\er
and
\br  \nonumber
\bar{\chi}_1 &=& -\frac{1}{2}  \bar{q} q X,\\ \nonumber
\bar{\chi}_2 &=& \frac{1}{4} i [2 q \pa_x \bar{q} X +  \bar{q}  q \pa_x X],\\
\label{chisd}
\bar{\chi}_3 &=& \frac{1}{8}  [3 q  \pa_x \bar{q} \pa_x X + 3 q  X ( q  \bar{q}^2 + \pa^2_x \bar{q}) +  \bar{q}  q \pa^2_x X],\\
\nonumber
\bar{\chi}_4 &=&-\frac{i}{16} [  q \bar{q}^2\( 5 q \pa_x X + 4 X \pa_x  q \) + 6  q  \pa^2_x \bar{q} \pa_x X + 4 q \pa_x \bar{q} \pa^2_x X + 4 q X \pa^3_x \bar{q} + \bar{q} q \( 16 X q \pa_x \bar{q} + \pa^3_x X \) ],\\
\\ 
\nonumber
\bar{\chi}_5 &=& -\frac{1}{32} q \Big\{ 10  \bar{q}^3 q^2 X + \bar{q}^2 \( 9 \pa_x X \pa_x  q + 7 q \pa^2_x X + 5 X \pa_x^2 q \) + \\
\nonumber
&& 5  [ 2 \pa^2_x \bar{q} \pa^2_x X + 2 \pa^3_x \bar{q} \pa_x X +  \pa_x^3 X \pa_x \bar{q} + X (5 q (\pa_x \bar{q})^2 + \pa^4_x \bar{q})]+ \\
\nonumber
&& \bar{q} [ 2 \pa_x \bar{q} (17 q \pa_x X + 15 X \pa_x  q ) + 30 X  q \pa^2_x \bar{q} + \pa^4_x X ] \Big\}. 
\er

\end{document}